\documentstyle[aps,prd,floats,graphicx]{revtex}

\def\fun#1#2{\lower3.6pt\vbox{\baselineskip0pt\lineskip.9pt
        \ialign{$\mathsurround=0pt#1\hfill##\hfil$\crcr#2\crcr\sim\crcr}}}

\renewcommand\({\left(}
\renewcommand\){\right)}
\renewcommand\[{\left[}
\renewcommand\]{\right]}

\newcommand\eq[1]{Eq.~(\ref{#1})}
\newcommand\eqs[2]{Eqs.~(\ref{#1}) and (\ref{#2})}

\newcommand\ee{\end{equation}}
\newcommand\be{\begin{equation}}
\newcommand\eea{\end{eqnarray}}
\newcommand\bea{\begin{eqnarray}}

\newcommand\GeV{\,\mbox{GeV}}

\newcommand\mpl{M_{\rm P}}

\newcommand\lsim{\mathrel{\rlap{\lower4pt\hbox{\hskip1pt$\sim$}}
    \raise1pt\hbox{$<$}}}
\newcommand\gsim{\mathrel{\rlap{\lower4pt\hbox{\hskip1pt$\sim$}}
    \raise1pt\hbox{$>$}}}

\newcommand\diff{\mbox d}

\def\dslash{\not{\hbox{\kern-2pt $\partial$}}}
\def\Dslash{\not{\hbox{\kern-4pt $D$}}}
\def\Oslash{\not{\hbox{\kern-4pt $O$}}}
\def\Qslash{\not{\hbox{\kern-4pt $Q$}}}
\def\pslash{\not{\hbox{\kern-2.3pt $p$}}}
\def\kslash{\not{\hbox{\kern-2.3pt $k$}}}
\def\qslash{\not{\hbox{\kern-2.3pt $q$}}}

 \newtoks\slashfraction
 \slashfraction={.13}
 \def\slash#1{\setbox0\hbox{$ #1 $}
 \setbox0\hbox to \the\slashfraction\wd0{\hss \box0}/\box0 }

\def\ee{\end{equation}}
\def\be{\begin{equation}}

\def\call{{\cal L}}

\def\calp{{\cal P}}

\newcommand\bfk{{\bf k}}

\newcommand\bfx{{\bf x}}

\newcommand\sub[1]{_{\rm #1}}

\begin{document}
\draft
 \twocolumn[\hsize\textwidth\columnwidth\hsize\csname
 @twocolumnfalse\endcsname

\preprint{}

\title{Generating the curvature perturbation without an inflaton}

\author{David H.~Lyth$^1$ and David Wands$^2$}

\address{$^1$\ Physics Department, Lancaster University, Lancaster LA1 4YB,
United Kingdom}

\address{$^2$\ Relativity and Cosmology Group, School of Computer Science
and Mathematics,\\University of Portsmouth, Portsmouth~PO1~2EG,
United Kingdom}

\date{\today}
\maketitle

\begin{abstract}
We present a mechanism for the origin of the large-scale curvature
perturbation in our Universe by the late decay of a massive scalar
field, the curvaton. The curvaton is light during a period of
cosmological inflation, when it acquires a perturbation with an
almost scale-invariant spectrum. This corresponds initially to  an
isocurvature density perturbation, which  generates the curvature
perturbation after inflation when the curvaton density becomes a
significant fraction of the total. The isocurvature density
perturbation disappears
if the curvaton completely decays into thermalised radiation. Any
residual isocurvature perturbation is 100\% correlated with the
curvature. The same mechanism can also generate the curvature
perturbation in pre big bang/ekpyrotic models, provided that the
curvaton has a suitable non-canonical kinetic term so as to
generate a flat spectrum.
\end{abstract}
\vskip 1pc \pacs{98.80.Cq \ \ 04.50.+h 
\hfill 
Physics Letters {\bf B524}, 5-14 (2002) \ \ PU-RCG-01/33 \ \ hep-ph/0110002v2}
%
 ]

\paragraph*{Introduction.}

It is now widely accepted that the dominant cause of structure in
the Universe is a spatial curvature perturbation \cite{book}. This
perturbation is present on cosmological scales  a few Hubble times
before these scales enter the horizon, at which stage it is
time-independent with an almost flat spectrum. One of the main
objectives of theoretical cosmology is to understand its origin.

The usual assumption is that  the curvature perturbation
originates during inflation, from the quantum  fluctuation of the
slowly-rolling inflaton field. As cosmological scales  leave the
horizon, the quantum fluctuation is converted to a classical
gaussian perturbation with an almost flat spectrum, generating
immediately the required curvature perturbation which is constant
until the approach of horizon entry. This idea has the advantage
that the prediction for the spectrum is independent of what goes
on between the end of inflation and horizon entry
\cite{bst,dhl,book}. The spectrum depends only on the form of the
potential and on the theory of gravity during inflation (usually
taken to be Einstein gravity), providing therefore a direct probe
of conditions during this era.  On the other hand, the demand that
inflation should produce the curvature perturbation in this
particular way is very restrictive, ruling out or disfavouring
several otherwise attractive models of inflation.

In this note we point out that the primordial curvature perturbation
may have a completely different origin, namely the quantum fluctuation
during inflation of a  light scalar field
which  is not the slowly-rolling inflaton, and need have
nothing to do with the fields driving of inflation.
We call this field the curvaton. The
curvaton creates the curvature perturbation in two separate stages.
First, its  quantum fluctuation during inflation is converted
at horizon exit to a classical perturbation with a flat spectrum.
Then, after inflation, the perturbation in the curvaton field is
converted into a curvature perturbation.
In contrast with the usual mechanism, the generation of curvature by
the curvaton requires
no assumption about  the nature of inflation,  beyond the
requirement that  (if the curvaton has a canonical kinetic term)
the Hubble parameter is practically constant.
Instead, it requires certain properties of the curvaton and of the cosmology
after inflation so that the required curvature perturbation will be
generated.
We shall explore the simplest setup, consisting of the following
sequence of events. First,
the curvaton field starts to oscillate during some
radiation-dominated era,  so that it
constitutes matter with an isocurvature density perturbation.
Second, the oscillation persists for many Hubble times so
that a significant curvature perturbation is generated.
Finally,  before neutrino decoupling, the curvaton
decays and the curvature perturbation remains constant until the approach
of horizon entry.

We shall show that under these conditions the quantum fluctuation of
the curvaton during inflation is converted into a curvature
perturbation after decay according to the formula
\be
\zeta
 \sim r \delta
\label{result}
 \label{1}
\ee
where $\delta$ is the isocurvature fractional density perturbation in
the curvaton before it decays and $r$ is the fraction of the final
radiation that the decay produces.

The mechanism that we are describing may succintly be described as the
conversion of an isocurvature perturbation into a curvature
perturbation.  It was actually discovered more than a decade ago by
Mollerach \cite{silvia}, who corrected the prevailing misconception
that no conversion would occur. At the time the conversion was
regarded as a negative feature, because the focus was on finding a
good mechanism for generating an isocurvature perturbation. For this
reason, and also because cosmology involving late-decaying scalars was
not considered to be very likely, the conversion mechanism has
received little attention\footnote{The possibility of converting an
  initially isocurvature perturbation into an adiabatic one was noted
  by Linde and Mukhanov\cite{andrei}}.  
The situation now is very different. On
the observational side, we know that the curvature perturbation
provides the principle origin of structure. On the theoretical side,
late-decaying scalars are routinely invoked in cosmology, and are
ubiquitous in extensions of the Standard Model of particle physics.
Also, one is now aware of inflation models whose only defect is their
failure to generate the curvature perturbation from the inflaton, some
of which will be mentioned later.

Before ending this introduction, we need to emphasise that the
curvaton  can produce a curvature perturbation {\em without} any
accompanying isocurvature perturbation at late times. This is the
reason why we include in our setup the requirement that the
curvaton decays before neutrino decoupling. If it decays later,
the curvature perturbation may be accompanied by a significant
isocurvature neutrino perturbation as discussed recently by Hu
\cite{wayne}.

\paragraph*{The curvature perturbation.}

The spatial curvature perturbation is of interest only on comoving
scales much bigger than the Hubble scale (super-horizon scales).
To define it one has to specify a foliation of spacetime into
spacelike hypersurfaces (slicing), and the most convenient choice
is the slicing of uniform energy density (or the slicing
orthogonal to comoving worldlines, which is practically the same
on super-horizon scales). The curvature perturbation on
uniform-density slices~\cite{bardeen88,ms,wmll} is given by the
metric perturbation $\zeta$, defined with a suitable coordinate
choice by the line element
\be
\diff \ell^2 = a^2(t) \( 1 +2\zeta
\) \diff x^i \diff x^j \,.
\ee
On cosmological scales, the spectrum $\calp_\zeta$ of $\zeta$ at
the approach of horizon entry is almost flat, with magnitude of
order $10^{-10}$.

The time-dependence of $\zeta$ on large scales is given by
\cite{wmll}
\be \dot\zeta = -\frac{H}{\rho + P} \delta P\sub{nad}
\label{zetadot} \,, \ee
where $H\equiv \dot a/a$ is the Hubble
parameter, $\rho$ is the energy density, $P$ is the pressure and
$\delta P\sub{nad}$ is the pressure perturbation on
uniform-density slices (the non-adiabatic pressure perturbation).

In the usual scenario where $\zeta$ is generated during inflation
through the perturbation of a single-component inflaton field, it
becomes practically time-independent soon after horizon exit and
remains so until the approach of horizon entry. The mechanism that
we are proposing starts instead with a negligible curvature
perturbation, which is generated later through a non-adiabatic
pressure perturbation associated with the curvaton perturbation.

\paragraph*{The curvaton field.}

The curvaton field $\sigma$ lives in an unperturbed Robertson-Walker
spacetime characterised by the line element
\be
\diff s^2 = \diff t^2 - a^2(t)\delta_{ij} \diff x^i \diff x^j
\ee
and its lagrangian is
\be
\call_\sigma = \frac12 \dot\sigma^2
- \frac12 \left(\nabla\sigma\right)^2 - V(\sigma)
\ee
The potential $V$ depends of course on all scalar fields
but we exhibit only the dependence on $\sigma$ which
is assumed to have no significant coupling to the fields driving
inflation.

The initial era for our discussion is the one which begins several
Hubble times before the observable Universe leaves the horizion
during an inflationary phase, and ends several Hubble times after
the smallest cosmological scale leaves the horizon. During this
era, we assume that the Hubble parameter $H\equiv \dot a/a$ is
almost constant, that is \be \epsilon_H\equiv -\dot H/H^2\ll 1 \,.
\label{doth} \ee In the usual slow-roll paradigm with Einstein
gravity, $2\epsilon_H\simeq(\mpl V'/V)^2$ where $V(\phi)$ is the
inflationary potential, $\phi$ is the slowly rolling inflaton and
$\mpl=2\times 10^{18}\GeV$ is the reduced Planck mass.  However
for our mechanism we need not assume any specific paradigm for
inflation.

We assume that the  curvature perturbations is  negligible during
inflation.  For slow-roll inflation with Einstein gravity this
requires $H \lsim 10^{-5}\epsilon_H^\frac12 \mpl $ or $V^\frac14\lsim
10^{-2}\epsilon_H^\frac14 \mpl$.  In any case, an inflation model with
Einstein gravity requires $H\lsim 10^{-5}\mpl$ from the cosmic
microwave background limit on gravitational waves.

We write at any given time
\be
\sigma(\bfx) = \sigma + \delta\sigma(\bfx)
\ee
(Throughout we adopt the convenient notation that the absence of an
argument denotes the unperturbed quantity.)
Like any cosmological quantity the spatial dependence of
$\delta\sigma$ can be Fourier-expanded in a comoving box much larger than
the observable Universe, but it is unnecessary and in fact undesirable
for the box to be indefinitely large.  Failure to limit the box size
leads among other things to an indefinitely large fluctuation for any
quantity with flat spectrum. It is a source of confusion in the usual
case of inflaton-generated curvature and the same would be true for
curvaton-generated curvature.

The unperturbed curvaton field satisfies \be \ddot\sigma
+3H\dot\sigma  + V_\sigma = 0 \,, \label{unpert} \ee where
$H\equiv \dot a/a$ is the Hubble parameter, and
a subscript $\sigma$ denotes $\partial /\partial  \sigma$.

We are interested in the perturbation $\delta\sigma_\bfk$,
 where $\bfk$ denotes the comoving
momentum. It is
 conveniently defined on the
spatially-flat slicing.
In general, the scalar field perturbations
on this slicing  satisfy to first order
the set of coupled equations~\cite{tn}
\bea
\ddot{\delta\phi}_i + 3H\dot{\delta\phi}_i + {k^2\over a^2}
\delta\phi_i
 \hspace*{1in} && \nonumber \\
+ \sum_i \left[ V_{\phi_i\phi_j} - {1\over\mpl^2a^3}
  \left( {a^3\over H}\dot\phi_i\dot\phi_j \right)^\cdot \right]
\delta\phi_j &=& 0 \eea For simplicity we assume that  $\sigma$ is
sufficiently decoupled from the other perturbations that the
latter can be ignored, leading to
\be \ddot{\delta\sigma}_\bfk +
3H\dot{\delta\sigma}_\bfk + \( (k/a)^2 + V_{\sigma\sigma} \)
\delta\sigma_\bfk = 0 \,. \label{perturb0}
\ee

We assume that the curvaton potential is sufficiently flat during
inflation,
\be
|V_{\sigma\sigma}|\ll H^2 \,, \label{flatness}
\ee
and that on cosmological scales each Fourier component is in the
vacuum state well before horizon exit. The vacuum fluctuation then
causes a classical perturbation $\delta \sigma_\bfk$ well after
horizon exit, which satisfies \eq{perturb0} with negligible
gradient term,
\be
\ddot{\delta\sigma}_\bfk +
3H\dot{\delta\sigma}_\bfk +  V_{\sigma\sigma} \delta\sigma_\bfk =
0 \,. \label{perturb}
\ee
The perturbation is gaussian, and in the
limit where \eq{flatness} is very well satisfied
 its spectrum given by
\be
\calp_{\delta \sigma}^\frac12 =\frac{H_*}{2\pi} \,.
\ee
The star denotes the epoch of horizon exit, $k=a_*H_*$, and, by
virtue of \eq{doth}, $\calp_\sigma$ is almost flat. To be more
precise, the spectral tilt of the perturbation is given by
\be
n_\sigma \equiv {d\ln\calp_\sigma \over d\ln k}
 = 2\frac{\dot H_*}{H_*^2} + {2\over3}{\(V_{\sigma\sigma}\)_*\over H_*^2} \,.
\label{nsigma} \ee


\paragraph*{Oscillating phase.}

We now move on to the epoch when the curvaton field starts to
oscillate around the minimum of its potential.
We suppose that the oscillation
starts during some radiation-dominated era. It may not be the one in
which nucleosynthesis occurs, but if it is we require that the
oscillation starts well before cosmological scales enter the horizon.
We assume that Einstein gravity is valid from the oscillation time
onwards, so that the total energy density is $\rho=3H^2\mpl^2$.

We assume that the curvaton continues to satisfy \eq{unpert}, and
that its perturbation continues to satisfy \eq{perturb}, after
inflation.
Assuming that the potential $V(\sigma)$ is quadratic, $V=m^2
\sigma^2/2$, then oscillations start at the epoch $H \sim m$.
Also, \eq{perturb} for $\sigma$ and \eq{unpert} for $\delta\sigma$
are then the same and the ratio $\delta\sigma/\sigma$ remains
fixed on super-horizon scales.  This assumption can easily be
relaxed, and in particular one can handle the situation where the
curvaton may initially be near a maximum of the potential
\cite{myaxion}. Any evolution of $\delta\sigma$ on super-horizon
scales leads to an overall scale-independent factor which will not
spoil the flatness of the spectrum.  Even if the potential is not
quadratic at the onset of oscillation, it will become practically
quadratic after a few Hubble times as the oscillation amplitude
decreases.

We are interested in the curvaton energy density
\be
\rho_\sigma(\bfx) = \rho_\sigma + \delta\rho_\sigma(\bfx)
\ee
and in the density contrast
\be
\delta\equiv
\frac{\delta\rho_\sigma}{\langle  \rho_\sigma \rangle} \ee
Since the spatial gradients are negligible on super-horizon
scales, the oscillation is harmonic at each point in space and
\be
\rho_\sigma(\bfx) =\frac12m^2\sigma^2(\bfx) \,,
\label{rhosigma}
\ee
where $\sigma(\bfx)$ is the amplitude of the oscillation.

When the oscillation starts, the mean-square perturbation of
$\sigma$ is given by
\be
\langle (\delta\sigma )^2 \rangle =
\int^{k\sub{max}}_{k\sub{min}} \calp_\sigma(k) \frac{\diff k}{k}
\,.
\ee
As discussed in \cite{myaxion}, the short distance cutoff at the
epoch when the oscillation starts is $k\sub{max}\sim (\tilde a
\tilde H)$ where a tilde denotes this epoch,  sub-horizon modes
having red-shifted away. Also,  since we are working in a box not
too much bigger than the the observable Universe the long distance
cutoff is $k\sub{min}\sim a_0  H_0$. Assuming that  $\calp_\sigma$
is flat this gives the estimate
\be
\frac{\langle (\delta\sigma
)^2 \rangle}{\sigma^2} = \left( \frac{H_*}{2\pi\sigma_*} \right)^2
\ln(k\sub{max}/k\sub{min})
  \sim (H_*/\sigma_*)^2
\,. \label{delsig}
\ee
(If $\calp_\sigma$ increases dramatically
on small scales, the estimate has to be increased appropriately.)

If  $H_*\ll \sigma_*$, the field perturbation is small and
\be
\delta = 2\frac{{\delta\sigma}}{\sigma}
\ee
This is a time-independent gaussian perturbation with a flat
spectrum given by
\be
\calp_\delta^\frac12= 2\calp_\sigma^\frac12/\sigma
=\frac{H_*}{\pi \sigma_* } \ll 1 \,. \label{gspec}
\ee

In the opposite regime, $H_*\gg \sigma_*$,
the perturbation is bigger than the unperturbed value and
\be
\delta=\frac{({\delta\sigma})^2} {\langle ({\delta\sigma})^2
\rangle} \,.
\ee
This is again time-independent, but is now the square of a
gaussian quantity (a $\chi^2$ quantity). Its spectrum is  flat up
to logarithims \cite{myaxion}
\be \calp_\delta(k) = 4\ln(k/k\sub{min})
\frac{\calp_\sigma(k\sub{min})^2} {\langle ({\delta\sigma})^2
\rangle} \sim 1 \label{nongspec} \,.
\ee

We shall show that the curvature perturbation is a multiple of
$\delta$, which means that it is gaussian in the regime $H_*\ll
\sigma_*$ but a $\chi^2$ non-gaussian quantity in the opposite
regime. A $\chi^2$ curvature perturbation is strongly forbidden by
observation~\cite{fffs}, which therefore requires
\be
H_* \ll
\sigma_* \label{hbound} \,.
\ee

\paragraph*{Generating the curvature perturbation}

Once the curvaton field starts to oscillate the energy density
becomes a mixture of matter (the curvaton) and radiation.
According to \eq{zetadot} the generation of the curvature
perturbation begins at that point, because the pressure
perturbation corresponding to this mixture is non-adiabatic. It
ends when the pressure perturbation again becomes adiabatic, which
is at the epoch of curvaton matter domination, or the epoch of
curvaton decay, whichever is earlier.

The curvature perturbation finally generated could be precisely calculated
from \eq{zetadot} knowing the decay rate $\Gamma$ of the curvaton,
but for an estimate it is enough to assume that the decay occurs
instantaneously at the epoch $H=\Gamma$. In that case one can
avoid the use of   \eq{zetadot} altogether by considering separately
the curvature perturbations $\zeta\sub r$ and $\zeta_\sigma$ on
respectively slices of uniform radiation- and matter density.
These are separately conserved \cite{wmll}
 as  the radiation and matter are perfect non-interacting fluids.
The curvature perturbations are given by \cite{wmll}
\bea
\zeta &=& -H\frac{\delta\rho}{\dot\rho}\\
\zeta\sub r &=&  -H\frac{\delta\rho\sub r }{\dot\rho\sub r}
=\frac14 \frac{\delta\rho\sub r}{\rho\sub r}\\
\zeta_\sigma &=&  -H\frac{\delta\rho_\sigma}{\dot\rho_\sigma}
=\frac13 \frac{\delta\rho_\sigma}{\rho_\sigma}\equiv \frac13\delta
\,,
\eea
where the density perturbations are defined on the flat slicing of spacetime.
(Note that the constancy of $\zeta_\sigma$ is equivalent to the constancy
of $\delta$  for the oscillating field which we noted earlier.)
Using these results the curvature perturbation is
\be
\zeta
= \frac{4\rho\sub r \zeta\sub r + 3\rho_\sigma \zeta_\sigma}
 {4\rho\sub r + 3\rho_\sigma} \,.
\label{weightedsum}
\ee

Before the oscillation begins, $\zeta=\zeta\sub r$ which we are supposing
is negligible. It follows that
\be
\zeta
= \frac{\rho_\sigma}
 {4 \rho\sub r + 3\rho_\sigma}  \delta
\,.
\ee
This calculation applies until the curvaton decays, after which
$\zeta$ is constant.
If the curvaton dominates the energy density before decay,
the final value of $\zeta$ is
\be
\zeta = \frac13 \delta
\,.
\ee
In the opposite case, the
curvaton  density just before decay is some fraction
 $r<1$ of the radiation density. Making the approximation
$r\ll 1$,
\be r \simeq {1\over6} \left( \frac{\sigma_*}{\mpl} \right)^2
\left( \frac{m}{\Gamma} \right)^{1/2} \,, \label{r} \ee
and the final curvature perturbation  is
\be \zeta = \frac14
r\delta \,.
\ee
Dropping the prefactors $\frac13$ and $\frac14$, the spectrum of
the curvature perturbation is given in the Gaussian regime,
$H_*\ll \sigma_*$, by
\be
\calp_\zeta^\frac12 \simeq r \frac{H_*}{\pi \sigma_* }
\label{specpred} \,.
\ee

The scale dependence of the spectrum is the same as that of
$\delta\sigma$. {}From \eqs{doth}{nsigma} its spectral index $n$
is given by
\be
n-1\equiv  \frac{\diff\ln\calp_\zeta}{\diff \ln k} =
2\frac{\dot H_*}{H_*^2} +\frac23 \frac{m^2}{H_*^2} \,.
\label{npred}
\ee
If we relax the demand that the potential is quadratic, $m^2$ is
replaced by the effective mass-squared
$m_*^2=(V_{\sigma\sigma})_*$ which can be either positive or
negative. The observational constraint
at 95\% confidence level is $n=0.93\pm0.13$ \cite{wtz}, which
requires
\be
|m_*^2|/H_*^2 \lsim 0.1 \,.
\label{mcon}
\ee

The completely non-gaussian regime $H_*\gg \sigma_*$ is strongly
forbidden by observation \cite{fffs}, but the intermediate regime
is allowed provided that the non-gaussian component is small.
{}From \eq{rhosigma}, the curvature perturbation in that case is
of the form \cite{myaxion}
\be
\zeta = \frac r4 \[
2\frac{\delta\sigma}{\sigma} + \( \frac{\delta\sigma}{\sigma} \)^2
\] \,, \ee with \be \calp_{\delta\sigma/\sigma}^\frac12 =
\frac{H_*}{2\pi\sigma_*}
\ee
and, since the first term dominates
\be
r\simeq 10^{-5}/\calp_{\delta\sigma/\sigma}^\frac12 \,.
\ee
%
{}Observational constraints on non-gaussianity can place strong
upper limits on the small ratio $H_*/\sigma_*$, which it would be
interesting to evaluate.

\paragraph*{An isocurvature density  perturbation?}

At the epoch when perturbations first become observable  as
cosmological scales approach the horizon, the curvature
perturbation seems to be the dominant cause of structure  but it
may not be the only one. In particular, there may be an
isocurvature perturbation (one present on slices of uniform total
energy density) in the density of one or more of the constituents
of the Universe relative (conventionally) to the photon density.
In the standard picture the constituents are the photon plus
 (i) the cold dark matter (ii) the baryons  and (iii) the
 three neutrinos of the Standard Model which have travelled freely
since they fell out of equilibrium shortly before nucleosynthesis.
There could be an isocurvature density perturbation in any or all
of these three components. As described for instance in
\cite{martinb,trotta,amendola}, these isocurvature perturbations
could be a significant fraction of the total as far as present
microwave background observations are concerned, though the PLANCK
satellite will rule out (or detect) them at something like the
10\% level if their spectrum is flat. Going beyond the standard
picture, the dark matter might have non-trivial properties and
there may be neutrinos or other free-streaming matter with a
non-thermal momentum distribution.

An isocurvature density perturbation  may originate as the quantum
fluctuation of a scalar field during inflation. However, such a
field cannot be the inflaton, and in the usual scenario where the
latter generates the curvature perturbation it is hard to see why
the effect of an  isocurvature perturbation should be big enough
to be observable. {\em A priori} one expects that either it will
be dominant, which is forbidden by observation, or else
negligible. In contrast, if the curvature perturbation is
generated by a curvaton field, that same field may also generate
an isocurvature perturbation. In other words, the isocurvature
density perturbation of the curvaton field might be converted into
a mixture of a curvature and a
correlated isocurvature perturbation.

A study of this possibility is outside the scope of the present paper,
but we offer some brief comments. Consider first the case that the
curvaton decays before neutrino decoupling.
In that case the pre-existing radiation is most likely in thermal
equilibrium, with the momentum distribution of each species
specified by the densities of the baryon number and the three
lepton numbers. In this situation, if  the decay radiation
thermalises with the existing radiation, and generates no baryon
or lepton number, the curvaton can only generate a dark matter
isocurvature perturbation if some small fraction decays into
decoupled dark matter with an exceedingly small branching ratio.
If on the other hand the decay radiation thermalises but does
possess some baryon or lepton number asymmetry, this can generate
a baryon or neutrino isocurvature perturbation, which might be big
enough to observe if the asymmetry is not too small.  An example
of the latter possibility is given in \cite{hitoshi}, which
however neglects the dominant curvature perturbation.

If the curvaton decays after neutrino decoupling, the pre-existing
radiation consists of photons and neutrinos which are now
decoupled. In that case the curvaton decay will  cause a
perturbation in the relative abundance of neutrinos and  photons,
no matter whether it decays to photons or to neutrinos. In  other
words  it will (conventionally) generate a neutrino isocurvature
perturbation. At least if the curvaton decays to neutrinos, this
isocurvature perturbation should be big enough to observe in the
forseeable future \cite{wayne}. (Curvaton decay after
nucleosynthesis can also significnatly alter the epoch of
matter-radiation equality, which again may be observable.)

A distinctive prediction of curvaton decay that generates both the
curvature perturbation and an isocurvature perturbation at late
times is that the two perturbations, arising from a single initial
curvaton perturbation, must be completely correlated. Current
microwave background data alone cannot rule out a significant
contribution from an isocurvature perturbation correlated with the
curvature perturbation~\cite{trotta,amendola}, but future data
will give much tighter constaints~\cite{martinb}.

\paragraph*{The curvaton as a flat direction}

We have still to consider the nature of the curvaton in the context of
particle physics. In particular, we did not examine the fundamental
assumption that the curvaton potential satisfies the flatness
requirement \eq{flatness}.

Let us first suppose that the curvaton is a generic field, running
over an indefinitely large range
 $\sigma >0$ with the lower end the fixed point of a symmetry.
The potential will be sufficiently flat only
over some range
$\sigma<\sigma\sub{max}$, beyond which it
 rises too steeply. If inflation lasts long enough
we may expect $\sigma_*$ in our location
to have equal probablity of being anywhere
in the range $0<\sigma_*<\sigma\sub{max}$,
leading to the rough estimate $\sigma_*\sim \sigma\sub{max}$.

For simplicity assume that the symmetry forbids odd powers
of $\sigma$ in the potential.
Adopting the usual paradigm of supersymmetry, the
renormalizable (quadratic and quartic) terms of the potential
can be eliminated at the level of global supersymmetry (the
 curvaton can be chosen as a flat direction in field space).
However, at the supergravity level
 the typical effective mass-squared  $m^2(t)$,
of a generic field in the early Universe is of order $\pm H^2$,
the true mass $m$ being relevant only after $H$ falls below
$m$. This is  marginally in conflict with the \eq{mcon}.

There will also be an infinite number of non-renormalizable terms,
of the form
$\lambda_d\Lambda^{4-d}\sigma^d$ with $d\geq 6$, where $\Lambda$
is the ultra-violet cutoff at or below the Planck scale.
The generic couplings are $\lambda_d\sim 1$, and taking
 $d=6$ the potential is sufficiently flat to satisfy \eq{flatness} only
in the regime $\sigma\lsim \sqrt{\Lambda H}$. In that case one expects
$\sigma_* \sim \sqrt{\Lambda H}$ and therefore from \eq{specpred}
\be
\calp_\zeta \sim r^2 \frac {H_*}{\Lambda} \,.
\ee
Alternatively, at least the first few non-renormalizable terms
might be supressed by a symmetry. In that case one might have
$\tilde \sigma \sim \mpl$ leading to the much smaller estimate
\be
\calp_\zeta \sim r^2 \frac{H_*^2}{\mpl^2} \,.
\ee
The actual value
will probably lie between these extremes, and one sees that the
observed value $\calp_\zeta\sim 10^{-10}$ can be achieved with
quite reasonable  values of the parameters. In contrast with the
case where the curvature is generated by the inflaton, the
prediction does not depend on the derivative of any potential and
is in that respect under better control.

There is clearly quite a bit of uncertainty in this generic case.
On the other hand, it has the advantage of occuring rather naturally
in the context of particle physics. A recent example, which actually
appeared when the present paper was almost written, is reference
\cite{hitoshi} in which $\sigma$ is the scalar super-partner
of a right-handed neutrino.
(Note though that this paper does not take account our
isocurvature-curvature conversion mechanism, supposing instead that
the decay of the sneutrino will set up only an isocurvature baryon
density perturbation.)

\paragraph*{The curvaton as a pseudo-goldstone boson}

To achieve better control of the curvaton mechanism, one can
suppose that $\sigma$ is a pseudo-goldstone boson, so that
$\sigma\to\sigma+$const under the action of some spontaneously
broken global symmetry. In the limit where this symmetry is exact
the potential would be exactly flat which makes the approximate
flatness we require technically natural \footnote{In a similar
way, it has been pointed out  \cite{cs} that making the inflaton a
pseudo-Goldstone boson will keep a hybrid inflation potential
sufficiently flat. The `natural inflation' proposal \cite{natural}
that the inflaton is a pseudo-Goldstone boson in a non-hybrid
model does not keep the inflaton potential flat, because both the
potential and its slope vanish in the limit of unbroken
symmetry.}. Also, in this case $\sigma$ runs over some  finite
range $0<\sigma<v$  with periodic boundary conditions. If
inflation lasts long enough, there is equal probability of finding
$\sigma_*$ anywhere in this range so one expects $\sigma_*\sim v$.
Even if it does not, no particular value is favoured and one has
the same expectation.

As a simple example consider a complex field $\Sigma$ with a
mexican-hat potential whose U(1) symmetry is broken by
non-renormalisable terms:
\be
V(\Sigma) = (|\Sigma|^2- v^2 )^2 + {1\over72\mpl^2} \left( 2v^6 -
  \Sigma^6 - (\Sigma^*)^6 \right) \,.
\ee Putting $\Sigma=ve^{i\sigma/v}$ this gives \be V(\sigma) =
{v^6\over36\mpl^2} \left[ 1- \cos\left({6\sigma\over v}\right)
\right] \,, \ee and hence \be \left| {d^2V\over d\sigma^2} \right|
\leq {v^4\over\mpl^2} \,. \ee In order for the  pseudo-goldstone
boson $\sigma$ to acquire a spectrum of classical perturbations on
super-horizon scales, while the radial field remains fixed in its
vacuum manifold, $|\Sigma|=v$, we require
\be {v^2\over\mpl} \ll H_* \ll v \,. \label{hinf}
\ee
This also ensures that variations in the local value of $\sigma$
has no effect on the background Hubble expansion.

After $N>v/H$ e-folds of inflation, the local value of the
$\sigma$ field in the vacuum manifold is effectively randomised,
so that we expect $\sigma_* \sim v$ and effective mass
\be
m_*^2\equiv (V_{\sigma\sigma})_* \sim \pm v^2/\mpl \label{mass}
\,.
\ee
For simplicity we assume that $\sigma_*$ is close to a
minimum so that the potential is quadratic.
%

The field begins to oscillate when $H\sim m_\sigma$ at which
time $\rho_{\rm rad}/\rho_\sigma\sim (v/\mpl)^2\ll1$.
Note that this is independent of the reheat temperature $T_{\rm rh}$.
The energy density of the oscillating $\sigma$ field comes to dominate
over the energy density of the radiation when
\be
H_{\rm eq} \sim {v^6 \over \mpl^5} \,.
\ee
On the other hand if we assume that $\sigma$ decays with only
gravitational strength interactions, then the time of decay is given
by
\be
H\sub{decay} \sim \Gamma \sim {m_\sigma^3 \over \mpl^2}
 \approx {v^6 \over \mpl^5} \,.
\ee
(That this is before nucleosynthesis requires that $v>10^{12}$GeV,
i.e., $m_\sigma>100$TeV.)
Thus we generally expect $\rho_\sigma\sim \rho_{\rm rad}$ at the time
of decay, and hence from \eqs{gspec}{specpred}
\be
\calp_\zeta
 \simeq \frac{H_*^2}{v^2}
 \,.
\ee

When the curvaton is a pseudo-goldstone boson, the flatness
requirement \eq{flatness} presents no problem since the flatness
of the potential is protected by the global symmetry.   The mass
$\sim v$ of the radial field is not protected and in a generic
supergravity theory one would expect that its effective value
during inflation would receive contributions of order $H_*$.
%
%
On the other hand, $\sigma_*$ in our part of the Universe is
equally likely to lie anywhere in the range $0<\sigma_* <v$ where
the upper limit is the effective value of $v$ during inflation.
Therefore, if $H_*$ is bigger than the true (vacuum) value of $v$,
the Gaussianity constraint $H_*\ll \sigma_*$ represents  a
significant restriction on the supergravity theory.

\paragraph*{Pre big bang and ekpyrotic scenarios.}

In some models of the very early universe, such as the pre big bang
scenario~\cite{pbb}, there is essentially no curvature perturbation
produced on large scales 
\cite{bggmv,lwc} and these
models have been largely discarded as possible sources for the origin
of large-scale structure.  Yet if they can produce an almost
scale-invariant spectrum of isocurvature perturbations then they may
generate a primordial adiabatic curvature perturbation on all scales
{\em after} 
the pre-big-bang phase.  Because the pre big bang ends with an
explosive gravitational production of particle on small scales at
energies approaching the Planck scale, the late production of entropy,
such as from decaying massive fields, may be necessary in any case to
avoid over-production of dangerous relics~\cite{blo}.

The evolution during a pre big bang~\cite{pbb} or
ekpyrotic~\cite{ek,ek2,py} phase is far from slow-roll in the Einstein
frame ($|\epsilon_H|=|\dot{H}/H^2|>1$) which leads to a steep blue
spectrum of curvature perturbations during 
collapse~\cite{ekdhl}.
All massless moduli fields minimally coupled in the Einstein frame
have blue spectra with spectral tilt $n_\sigma=3$ in the
pre big bang or $n_\sigma=2$ in the ekpyrotic scenario~\cite{Wands}.
However it has previously been shown that a scale-invariant spectrum
may be generated in axion fields in the pre big bang
scenario~\cite{cew}.  Axion-type fields have a non-minimal kinetic
coupling to the dilaton field which yields a range of different scale
dependences being determined by the symmetries of the sigma-model
effective action and arbitrary constants of
integration~\cite{cew,axionspectra}. If the axion remains decoupled
during the uncertain transition from pre to post big bang era then
these large-scale perturbations remain isocurvature perturbations at
the start of the radiation dominated era~\cite{buonanno,darkaxions}.

Previous attempts to model structure formation in the pre big bang
have assumed that these axions remain decoupled and effectively
massless, only generating curvature perturbations when they re-enter
the horizon~\cite{seedaxions}.  This latter model gives distinctive
predictions for the spectrum of cmb anisotropies~\cite{seedcmb}, but
may be hard to reconcile with the latest observational data.  However
we have shown that an initial spectrum of axion perturbations can in
fact generate curvature perturbations on super-horizon scales
\footnote{Note added: This idea has also been studied in the context
  of pre big bang scenario by Enqvist and Sloth~\cite{es} in a
  preprint that appeared while this work was being written up.},
as has previously been suggested in Refs.~\cite{lwc,buonanno}.  If the
axions acquire a non-perturbative mass and come to contribute a
significant fraction to the total energy density before they decay in
the early universe, then they can act as a curvaton field and generate
a large-scale curvature perturbation long before horizon entry,
indistinguishable from that produced in a conventional inflation
model.

\paragraph*{Conclusions.}

We have drawn attention to the fact that the curvature perturbation
in the Universe need not be generated by the quantum fluctuation of
a slowly-rolling inflaton field, as is generally supposed.
Instead it may be generated by the quantum fluctuation of
a field that has nothing to do with the inflation model,
which we have called the curvaton.

The curvaton mechanism for the generation of curvature
perturbations makes it much easier to construct a viable model of
inflation. Inflation need not be of the slow-roll variety, and
even if it is there is no need for highly accurate slow-roll
demanded by the observed spectral index $|n-1|\lsim 1$. For
instance, extended inflation \cite{extended} was ruled out by the
predicted spectral index $n\lsim 0.7$ by the first COBE result
\cite{ll1} assuming the standard inflaton mechanism of curvature
generation, but it becomes viable with the curvaton mechanism. A
similar remark applies to modular inflation \cite{bdm}, which (at
least typically) also predicts too low a spectral index
\cite{covilyth3}. Another theoretically attractive model giving a
too low spectral index is described in \cite{cs}. One might even
have inflation where the inflaton is not rolling at all, such as
thermal inflation \cite{thermal}, though more than one bout of
such inflation would be necessary since at least in the usual
context of gravity-mediated supersymmetry breaking a single bout
gives only of order $10$ $e$-folds of inflation.


\paragraph*{Acknowledgements.}
The authors are grateful to Lev Kofman, Andrew Liddle, Andrei Linde
and Neil Turok for useful comments.
DW is supported by the Royal Society.

\newcommand\plpl[3]{Phys.\ Lett.\ {\bf #1}  (#3) #2}
\newcommand\np[3]{Nucl.\ Phys.\ {\bf #1}  (#3) #2}
\newcommand\pr[3]{Phys.\ Rep.\ {\bf #1}  (#3) #2}
\newcommand\prlprl[3]{Phys.\ Rev.\ Lett.\ {\bf #1}  (#3)  #2}
\newcommand\prdprd[3]{Phys.\ Rev.\ D{\bf #1}  (#3) #2}
\newcommand\ptp[3]{Prog.\ Theor.\ Phys.\ {\bf #1}  (#3)  #2 }
\newcommand\rpp[3]{Rep.\ on Prog.\ in Phys.\ {\bf #1} (#3) #2}
\newcommand\jhep[2]{JHEP #1 (#2)}
\newcommand\grg[3]{Gen.\ Rel.\ Grav.\ {\bf #1}  (#3) #2}
\newcommand\mnras[3]{MNRAS {\bf #1}   (#3) #2}
\newcommand\apjl[3]{Astrophys.\ J.\ Lett.\ {\bf #1}  (#3) #2}

\end{document}